\documentclass[]{elsarticle}
\usepackage{lineno,hyperref}
\usepackage{epstopdf} 
\usepackage{color}
\usepackage{lscape}
\modulolinenumbers[5]
\biboptions{sort&compress}
\usepackage{graphicx}
\usepackage{amssymb}
\begin{document}
\begin{frontmatter}
\title{Relationship between the shear modulus and volume relaxation in high-entropy metallic glasses: experiment and physical origin}

\author[STANKIN]{R.S. Khmyrov}
\author[VSPU]{A.S. Makarov}
\author[NWPU]{J.C. Qiao}
\author[IFTT]{N.P. Kobelev}
\author[VSPU]{V.A. Khonik}
\cortext[cor]{Corresponding author} 
\ead{v.a.khonik@yandex.ru} 

\address[STANKIN] {Moscow State University of Technology "STANKIN", Moscow 127055, Russia}
\address[VSPU] {Department of General Physics, Voronezh State Pedagogical
University,  Voronezh 394043, Russia}
\address[NWPU] {School of Mechanics, Civil Engineering and Architecture, Northwestern Polytechnical University, Xi’an 710072, China}
\address[IFTT] {Institute of Solid State Physics RAS, Chernogolovka 142432, Russia}

\begin{abstract}
	We performed parallel measurements of the high-frequency shear modulus $G$ and relative volume  $\Delta V/V$  for high-entropy  Ti$_{20}$Zr$_{20}$Hf$_{20}$Be$_{20}$Cu$_{20}$  and Ti$_{20}$Zr$_{20}$Hf$_{20}$Be$_{20}$Ni$_{20}$ glasses upon heating from room temperature up to the complete crystallization. The changes of these properties due to structural relaxation both below and above the glass transition temperature are singled out. It is shown  that these changes for both initial and preannealed samples can be  well described within the framework of the Interstitialcy theory. It is found that the whole relaxation process in the full temperature range  of the experiments for both samples' states can be characterized by a single dimensionless temperature-independent parameter $K_i=dln\;G/dln\;V$, which equals to -44 and -53 for the above glasses, respectively, and strongly points at interstitial-type defects as a source of relaxation. We also show that the relaxation of the relative volume linearly depends on the defect concentration. This behavior can be described by another dimensionless temperature-independent parameter, which is related with the relaxation volume of  defects. Possible contribution of vacancy-like defects into the relaxation is discussed.

\end{abstract}


\end{frontmatter}

\section{Introduction}

It has long been known that the specific volume of metallic glasses (MGs) is normally by about 0.5\% larger than  that of crystalline counterparts \cite{ChenRepProgPhys1980}. This excess volume is changed upon any heat treatment due to spontaneous atomic rearrangements generally called structural relaxation. Although  volume alteration  associated with structural relaxation is small, it leads to appreciable changes of most of MGs' physical properties \cite{ChenRepProgPhys1980}. In particular, structural relaxation notably changes the elastic constants of MGs, especially the shear modulus $G$ \cite{ChenRepProgPhys1980}. This physical quantity is of particular interest since, first, it actually represents a thermodynamic parameter  because of being a 2$^{nd}$ derivative of the Helmholtz free energy with respect to the shear strain  \cite{Hirth,GranatoPRL1992}. Second, it constitutes a key physical parameter of the so-called "elastic models" and a few other theories on relaxation in glasses \cite{GranatoPRL1992,DyreRevModPhys2006}. 

One can consider that  alteration of $G$  upon structural relaxation is related with specific volume changes or, in other words, with the variation of the "free volume". Other quite different approaches to this issue are also possible \cite{MakarovJNCS2019}. Thus, it is important to understand how and why changes of the shear modulus depend on volume variations upon structural relaxation. 

The first experimental information on this issue was presented by Chen in 1978 \cite{ChenJAP1978} who showed for Pd-, Ni- and Fe-based  MGs that the derivative $-dln\;E/dln\;V=$ 15 to 17, where $E$ is the Young's modulus (this derivative is close to the derivative $dln\;G/dln\;V$, which we denote as $K$ hereafter). Much later Harms \textit{et al}. \cite{HarmsJNCS2003} found for bulk glassy Pd$_{40}$Cu$_{30}$Ni$_{10}$P$_{20}$ that $K=dln\;G/dln\;V=\frac{dG}{G}/\frac{dV}{V}\approx -25$ (that corresponds to  $dln\;E/dln\;V \approx -23$). 

A question arises what  these large values of the parameter $K$ mean.  A common approach to the structure of MGs assumes that it can be represented as a non-crystalline matrix with dominant structural packing and regions with low point symmetry commonly called defects \cite{ChengProgMaterSci2011}. These defects are considered  from very different viewpoints (e.g. Refs \cite{SpaepenActaMetall1977,ArgonActaMetall1979,vdBeukelActaMetMater1990,
EgamiProgMaterSci2011,LiuSciRep2017,KobelevUFN2023}) and structural relaxation is most often interpreted in terms of  changes of their concentration. However, different types of defects should lead to significantly different variations of the shear modulus  upon relaxation-induced changes of the volume, i.e. to different $K$-values. In particular, the free volume notions \cite{SpaepenActaMetall1977,ArgonActaMetall1979,vdBeukelActaMetMater1990}, despite of extensive critics of this approach (e.g \cite{MiracleMRSBull2007,ChengApplPhysLett2008}), 
 are  often used for the interpretation of volume changes upon structural relaxation \cite{HaruyamaApplPhysLett2006,HaruyamaActaMater2010,EvensonActaMater2011,BunzJAP2013}. However, the free volume in the form of vacancy-type defects would lead to a weak dependence of the shear modulus on the defect concentration  because these defects are expected to be spherically nearly symmetric and, therefore, should be  mostly unaffected by the external shear stress \cite{MakarovJPCM2020}. As a result, the quantity $K$  should be about $-3$ \cite{HarmsJNCS2003,MakarovJPCM2020}. Instead, the defects similar to dumbbell interstitials should be strongly anisotropic and constitute elastic dipoles, which are  markedly sensitive to the applied shear  \cite{GranatoPRL1992,DederichsJNuclMater1978,MakarovJPCM2020}.  This large sensitivity provides substantial anelastic shear strain and, therefore,  generation of these defects leads to a significant decrease of the shear modulus and, respectively, the parameter $K$ should be about $-30$  \cite{HarmsJNCS2003,MakarovJPCM2020}. This phenomenon is known as the diaelastic effect and it was first documented for crystals long ago \cite{HolderPhysRevLett1974,RobrockRadEff1975}. Thus, $K$-values are indicative  of the type of defects responsible for relaxation, as first suggested in Ref.\cite{MakarovJPCM2020}.

In the previous investigation, we performed parallel study of shear modulus and volume changes occurring upon \textit{isothermal} structural relaxation of a number of conventional Zr-based MGs and found the parameter $K$ to be -45 to -38 \cite{MakarovJPCM2020}. These $K$-values imply that structural relaxation is related to the evolution of interstitial-type defect system, as suggested by the Interstitialcy theory \cite{GranatoPRL1992,KobelevUFN2023}. 

In the present work, we focus on high-entropy metallic glasses, which have a high mixing entropy $\Delta S_{mix}=-R\sum\limits_{i=1}^nc_ilnc_i\geq{1.5} R$, where   $R$ is the universal gas constant, $c_i$ is the molar fraction of the \textit{i}-th element and $n$ is the number of constituent elements \cite{DuEncMater2022}. 
It was shown in recent years that these glasses have a number of interesting properties \cite{JiangNatCommun2021,DuanPRL2022,LuanNatComm2022,DuanPRL2024,AfoninAPL2024}. The goal of the present work is to understand whether and how the entropy state of glass influences its defect system \cite{AfoninAPL2024}. We performed parallel \textit{non-isothermal} (linear heating) shear modulus and  volume (length) measurements and found that the diaelastic effect in high-entropy glasses can be even larger than that in conventional MGs. Besides that, \textit{i}) we confirmed earlier conclusion on the origin of shear modulus vs volume relaxation, \textit{ii})  derived  a new dimensionless temperature-independent parameter related to the defect system and  \textit{iii})  designed relaxation master curves valid for both initial and preannealed samples in the whole temperature range under study (room temperature to the crystallization temperature).
   
\section{Experimental} 
 
High-entropy  Ti$_{20}$Zr$_{20}$Hf$_{20}$Be$_{20}$Cu$_{20}$  and Ti$_{20}$Zr$_{20}$Hf$_{20}$Be$_{20}$Ni$_{20}$ glasses (at.\%, TiZrHfBeCu and TiZrHfBeNi hereafter) (the mixing entropy $\Delta S_{mix}/R=1.61$) were prepared by melt suction and  X-ray confirmed to be completely amorphous. Measurements of the shear modulus were performed using the electromagnetic acoustic transformation method (EMAT) \cite{Vasil’evUFN1983}. This method is based on Lorentz interaction between the current induced in sample’s surface layer by an exciting coil and external magnetic field. By  scanning of current frequency one can excite transverse resonant vibrations  ($f\approx 400$ kHz to 600 kHz) of a sample (5$\times $5$\times $2 mm$^3$) using a pick-up coil. Frequency scanning was automatically performed every 10--15 s upon heating and the shear modulus was then calculated as $G(T) = G_{rt}f^2(T)/f^2_{rt}$, where $f_{rt}$ is the resonant frequency at room temperature and $G_{rt}$ is the room-temperature shear modulus. The error in the measurements of $G(T)$-changes  was estimated to be $\approx 5$ ppm near room temperature and about 100 ppm near the glass transition temperature $T_g$. Measurements were carried out in a vacuum of about 0.013 Pa. 

Dilatometric studies were performed by a Netzsch Dil 402 instrument on $2\times 2\times 20$ mm$^3$ specimens in flowing Ar atmosphere. Volume changes upon heating were taken  to be $\Delta V(T)/V_{rt}=3\Delta l(T)/l_{rt}$, where $V$ and $l$ are sample's volume and length, respectively, and   \textit{rt} refers to room  temperature hereafter. Calorimetric measurements were performed using a Hitachi DSC 7020 instrument operating in high purity (99.999\%) nitrogen atmosphere. Glass transition temperature $T_g$ and crystallization onset temperature $T_c$ were accepted as temperatures corresponding to the beginning of endothermal and exothermal reactions, respectively. All experiments were performed at a heating rate of 3 K/min. The densities of glassy TiZrHfBeCu and TiZrHfBeNi were found by the hydrostatic weighing to be 7470 kg/m$^3$ and 7460 kg/m$^3$, respectively.

\section{Results} 

\begin{figure}[t]
\begin{center}
\includegraphics[scale=0.7]{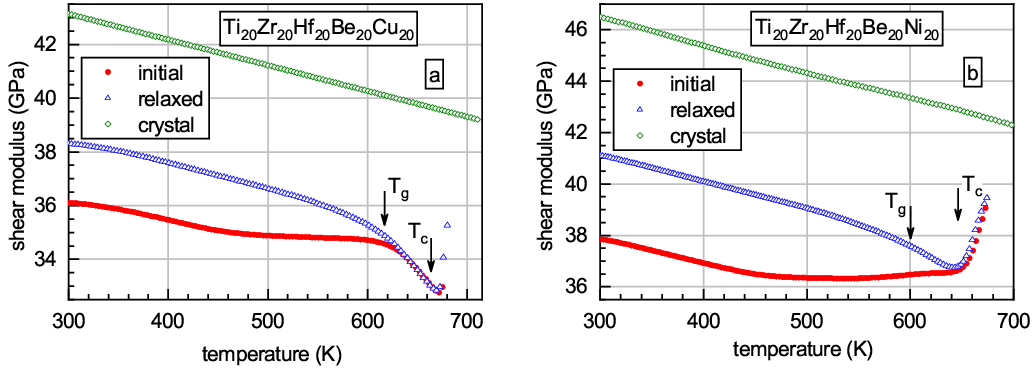}
\caption[*]{\label{Fig1.eps} Temperature dependences of the shear modulus of glassy TiZrHfBeCu (a) and  TiZrHfBeNi (b)  in the initial, relaxed and crystalline states measured at a rate of 3 K/min. Glass transition and crystallization onset temperatures, $T_g$ and $T_c$, are shown by the arrows. }
\end{center}
\end{figure} 

Figure \ref{Fig1.eps} shows temperature dependences of the shear modulus $G$ of the  MGs under investigation in the initial state, after relaxation performed by heating  into  the supercooled liquid state (up to 645 K and 625 K for TiZrHfBeCu and  TiZrHfBeNi, respectively) and after full crystallization obtained by heating up to 750 K. In general, $G(T)$- dependences for both MGs are similar although differ significantly  in details. Exothermal structural  relaxation in both cases starts at about 450 K -- 460 K and is manifested as an increase of the shear modulus over pure anharmonic decrease of $G$. For TiZrHfBeCu glass (Fig.\ref{Fig1.eps}a), entering the supercooled liquid state (i.e. above calorimetric $T_g$) results in a sharp increase of the slope $\vert dG/dT\vert$ but in the case of TiZrHfBeNi (Fig.\ref{Fig1.eps}b) this effect is not observed. It is to be noted that structural relaxation below $T_g$ (rise of the shear modulus) and the glass transition effect ($G$-decrease in the supercooled liquid state) in glassy TiZrHfBeCu (Fig.\ref{Fig1.eps}a)  are much more pronounced. In the relaxed state, the room-temperature shear modulus in both glasses is increased by $\approx 6\%$. Besides that, the upward $G$-change due to structural relaxation below $T_g$ in relaxed samples is absent. In all cases, crystallization leads to a significant increase of the shear modulus. In the crystalline state, $G$ decreases with temperature almost linearly and does not reveal any specific features.     

\begin{figure}[t]
\begin{center}
\includegraphics[scale=0.68]{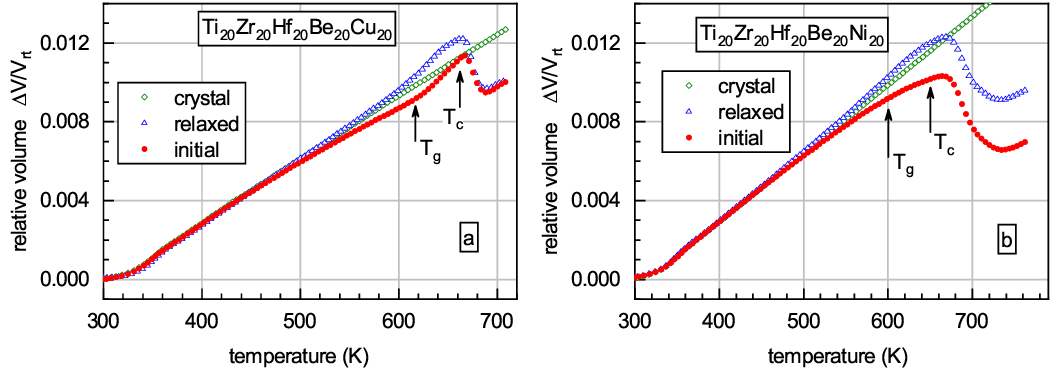}
\caption[*]{\label{Fig2.eps} Volume heat expansion of glassy TiZrHfBeCu  (a) and TiZrHfBeNi (b) in the initial, relaxed and crystalline states measured at a rate of 3 K/min. Glass transition and crystallization onset temperatures, $T_g$ and $T_c$, are shown by the arrows. }
\end{center}
\end{figure} 

The results of dilatometric measurements are  shown in Fig.\ref{Fig2.eps} in terms of temperature dependences of the volume changes presented as $\Delta V(T)/V_{rt}$ where $\Delta V(T)=V(T)-V_{rt}$. Again, the dilatometric curves for both MGs are similar but notably differ in details. Exothermal structural relaxation in the initial state is manifested as a downward change of $\Delta V/V_{rt}$, which starts at about 540 K to 560 K. Reaching of $T_g$ leads to a significant increase of the slope $d\Delta V/dT$ for      TiZrHfBeCu (Fig.\ref{Fig2.eps}a) while this effect is not observed in the case of TiZrHfBeNi (Fig.\ref{Fig2.eps}b). Further heating of both glasses results in volume contraction near calorimetric $T_c$. Relaxed samples subjected to preheating treatment  into the supercooled liquid state as described above do not show any relaxation-induced volume contraction below $T_g$ but reveal upward $\Delta V/V_{rt}$ change near and above $T_g$. Crystallization onset in both cases leads to volume contraction. Thermal expansion in the crystalline state is almost linear in temperature and does not show any peculiarities.   

Let us now relate the shear modulus and volume relaxation behavior presented  above. 

\section{Discussion} 
\subsection{Temperature dependence of density changes}

Quite a few relaxation phenomena in MGs can be quantitatively understood within the framework of the Interstitialcy theory (IT) originally suggested by Granato \cite{GranatoPRL1992}. The IT argues that melting of metals is related to a rapid increase of the concentration of dumbbell  interstitials, which remain identifiable structural units in the melt. Rapid melt quenching leads to the formation of solid glass, which contains a system of interstitial-type defects inherited from the melt. Thermoactivated change of the concentration of these defects is assumed responsible for the alteration of MGs' physical properties both below and above $T_g$. A detailed review of the IT approach is presented in Ref.\cite{KobelevUFN2023}.

The basic equation of the IT relates changes of the unrelaxed (high-frequency) shear modulus of glass $G$ with the concentration $c$ of frozen-in interstitial-type defects as 
\begin{equation}
G=\mu \;exp(-\alpha_g\beta_i c), \label{G}
\end{equation}
where $\mu$ is the shear modulus of the maternal crystal\footnote{By a maternal crystal we understand the polycrystal obtained by the complete crystallization of glass, which does not undergo any further phase transformations.}, $\alpha_g$ is a constant of the order of unity related to defects' stress field and $\beta_i$ is a dimensionless shear susceptibility (of about 20), which constitutes an integral parameter relating the defect-induced shear softening, heat effects, anharmonicity of interatomic interaction and defect structure of MGs \cite{KobelevUFN2023}. It is seen that the above equation describes a strong defect-induced diaelastic effect (shear softening due to an increase of the defect concentration). 

Defect formation  leads to volume changes. Considering that a  concentration $\Delta c$ of interstitial-type defects is created, one can calculate the relative volume changes as 

\begin{equation}
\left(\frac{\Delta V}{V}\right)_i=(r_i-1)\Delta c, \label{DeltaV_V} 
\end{equation}
where $r_i$ is the so-called relaxation volume for an interstial-type defect \cite{MakarovJPCM2020,MakarovJNCS2019}. Then, using the above basic IT equation, one can derive a relationship for a change of the defect concentration $\Delta c$ as a function of temperature upon linear heating using temperature depedences of the shear moduli of glass and maternal crystal together with room temperature  values of these quantities as
\begin{equation}
 \Delta c(T)=c(T)-c_{rt}=\frac{1}{\beta_i}ln\left[\frac{G_{rt}}{\mu_{rt}}\frac{\mu(T)}{G(T)}\right]. \label{Deltac}
 \end{equation}  
Then, taking into account the relation between the volume and corresponding density changes, $\Delta V/V=-\Delta \rho/\rho$, together with Eq.(\ref{DeltaV_V}) one arrives at the corresponding relaxation density change in the form
\begin{equation}
  \frac{\Delta \rho (T)}{\rho_{rt}}=\frac{r_i-1}{\beta_i}ln\left[\frac{\mu_{rt}}{G_{rt}}\frac{G(T)}{\mu(T)}\right]. \label{DeltaRho_Rho}
\end{equation}  
This equation shows that the relative density change as a function of temperature can be calculated using the data on the shear modulus given in Fig.\ref{Fig1.eps}. In addition, Eq.(\ref{DeltaRho_Rho}) can be used both for initial and relaxed states of glass (in the latter case, one should accept $G=G_{rel}$).

The corresponding experimental density changes can be determined using the data given in Fig.\ref{Fig2.eps} as  
\begin{equation}
\frac{\Delta \rho(T)}{\rho_{rt}}=\left[\frac{\Delta \rho (T)}{\rho_{rt}}\right]_{gl}-\left[\frac{\Delta \rho (T)}{\rho_{rt}}\right]_{cr}=\left[\frac{\Delta V (T)}{V_{rt}}\right]_{cr}-\left[\frac{\Delta V (T)}{V_{rt}}\right]_{gl}, \label{DeltaRhoExp} 
\end{equation}
where  the subscripts \textit{gl} and \textit{cr} refer to the glass and maternal crystal, respectively. 

For density calculation with Eq.(\ref{DeltaRho_Rho}), besides temperature dependences $G(T)$ and $\mu(T)$, one needs to know the shear susceptibility $\beta_i$ and the relaxation volume $r_i$ for an interstitial-type defect. The shear susceptibility was calculated according to Ref.\cite{AfoninJNCS2017} as
 $\beta_i=\frac{G_{rt}^{rel}-G_{rt}^{ini}}{\rho Q_{rel}}$, where $G_{rt}^{ini}$ and $G_{rt}^{rel}$ are the shear moduli of glass at room temperature in the initial and relaxed states, respectively, $\rho$ is the density and $Q_{rel}$ is the heat of structural relaxation, which is released during  warming up of initial sample to the supercooled liquid state as described above and cooling back to room temperature. This quantity was found to be 14600 J/kg and 18900 J/kg for TiZrHfBeCu and TiZrHfBeNi, respectively. With these $Q_{rel}$-values, the shear susceptibilities of these glasses are equal to 20 and 23, correspondingly. The relaxation volumes were accepted to be $r_i=1.45$ and $r_i=1.43$ for the above glasses (see below for details).  

\begin{figure}[t]
\begin{center}
\includegraphics[scale=0.68]{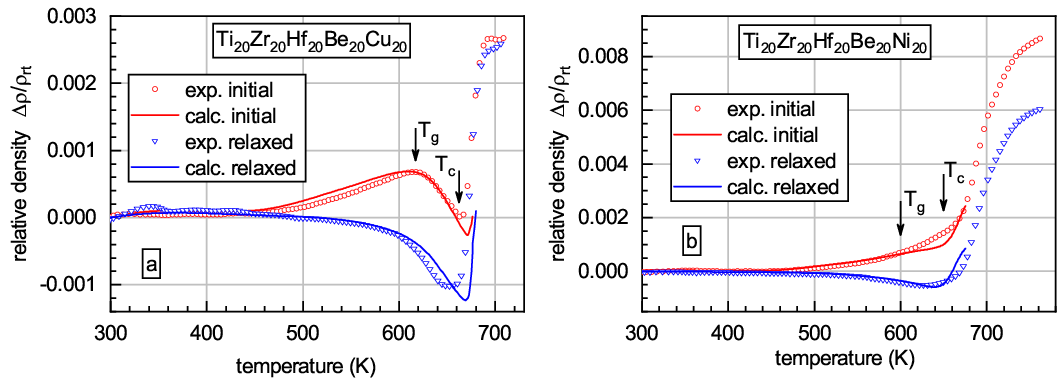}
\caption[*]{\label{Fig3.eps} Experimental relative density changes as a function of temperature of glassy TiZrHfBeCu  (a) and TiZrHfBeNi (b) in the initial and relaxed states derived using Eq.(\ref{DeltaRhoExp}) together with corresponding model data calculated with Eq.(\ref{DeltaRho_Rho}). It is seen that the calculation reproduces experimental data quite well. }
\end{center}
\end{figure} 

Figure \ref{Fig3.eps} shows experimental data on the relative density change for both MGs under study in the initial and relaxed stated replotted according to Eq.(\ref{DeltaRhoExp}) together with IT-based model calculation using Eq.(\ref{DeltaRho_Rho}) with the parameters indicated above. It is seen that the calculation provides a good reproduction of the experimental data. Meanwhile, the behavior of TiZrHfBeCu and TiZrHfBeNi  in the supercooled liquid state and upon crystallization is quite different. Indeed, \textit{i}) TiZrHfBeCu glass in the initial state displays much larger relaxation decrease of the density in the supercooled liquid state before crystallization onset, \textit{ii}) structural relaxation-induced rise of the density in this glass below $T_g$ is also much larger and \textit{iii}) crystallization-induced density increase is significantly larger as well. It is to be noted in this connection that TiZrHfBeCu and TiZrHfBeNi glasses display significantly different low temperature internal friction peaks, which are assumed to be a manifestation of  interstitial-like defect system \cite{AfoninJALCOM2024} and, besides that, have considerably different excess entropies \cite{AfoninAPL2024}. However, a good reproduction of the experimental data shown in Fig.\ref{Fig3.eps} within the IT-based approach suggests that all aforementioned  regularities in high-entropy MGs  are determined by the interrelation between shear modulus and volume relaxation, which is primarily governed by the alteration of the defect system as described by Eq.(\ref{Deltac}).

\subsection{Relaxation master curves }

On the other hand, shear modulus vs volume relaxation can be characterized using a single dimensionless parameter $K=\frac{dln\;G}{dln\;V}$ introduced in Ref.\cite{MakarovJPCM2020} as mentioned above in Introduction. This parameter indicates the origin of point(-like) defects responsible for relaxation both in crystalline and glassy states. For dumbbell interstitials this parameter becomes \cite{MakarovJPCM2020}  
\begin{equation}
  K_i=-\frac{\beta_i}{r_i-1}, \label{K}
  \end{equation}  
where the parameters $\beta_i$ and $r_i$ are described above.  Taking the aforementioned values $\beta_i=20$ and $\beta_i=23$, $r_i=1.45$ and $r_i=1.43$ for TiZrHfBeCu and TiZrHfBeNi, respectively, using Eq.(\ref{K}) one arrives at $K_i=-44$ and $K_i=-53$ for these glasses, correspondingly.

On the other hand, the parameter $K_i$ can be determined using Eq.(\ref{DeltaRho_Rho}). Let us rewrite this equation as
\begin{equation}
\delta (T)=K_i\Omega (T), \label{delta}
\end{equation}
where temperature-dependent dimensionless functions $\delta$ and $\Omega$ are defined as 
\begin{equation}
\delta (T)= ln\left[\frac{\mu_{rt}}{G_{rt}}\frac{G(T)}{\mu (T)}\right] \label{delta_mu}
\end{equation}
and 
\begin{equation}
\Omega (T)=\left[\frac{\Delta V(T)}{V_{rt}}\right]_{gl}-\left[\frac{\Delta V(T)}{V_{rt}}\right]_{cr}. \label{Omega}
\end{equation}
Equation (\ref{delta}) shows that the function $\delta (T)$ should be linear in the variable $\Omega (T)$ with a slope equal to a single temperature-independent parameter $K_i$ determined by Eq.(\ref{K}). 

\begin{figure}[t]
\begin{center}
\includegraphics[scale=0.68]{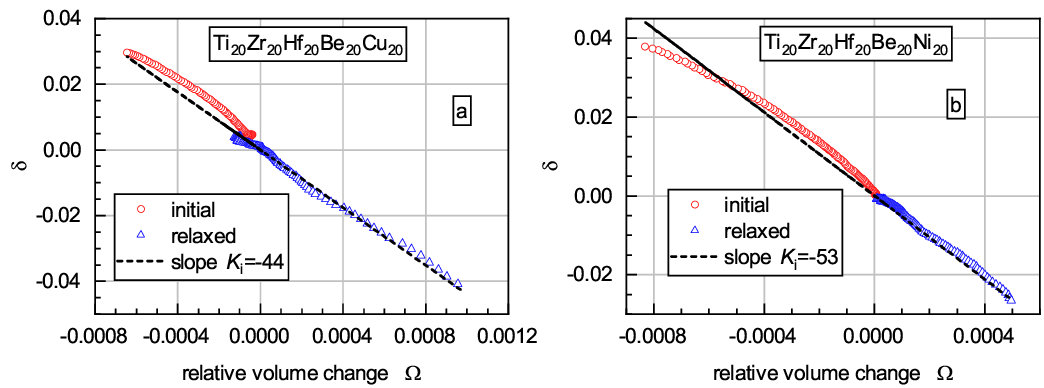}
\caption[*]{\label{Fig4.eps} Parameter $\delta $ characterizing temperature dependences of the shear moduli $G$ and $\mu$ and defined by Eq.(\ref{delta_mu}) as a function of the relative volume change $\Omega $ determined by Eq.(\ref{Omega}). The data  for glassy TiZrHfBeCu  (a) and TiZrHfBeNi (b) in the initial and relaxed states are shown. The dashed lines give the least square fits. It is seen that $\delta $ almost linearly depends on $\Omega $, in accordance with Eq.(\ref{delta}). The slopes equal to the parameter  $K_i=d\delta/d\Omega$ are indicated.  }
\end{center}
\end{figure}

Since all parameters and temperature-dependent functions are known, the dependence of $\delta(T)$ on $\Omega (T)$ can be plotted. This is done in Fig.\ref{Fig4.eps} for MGs under investigation in the initial and relaxed states. It is seen that $\delta$ indeed almost linearly decreases with $\Omega$ for both MGs under study. Meanwhile, in each case this linear dependence consists of two parts corresponding to the initial (red circles) and relaxed (blue triangles) states. This means that the experiments on initial and relaxed samples agree with each other within the model IT-based approach under consideration. The slopes $K_i=d\delta /d\Omega$ of the linear fits equal to $-44$ and $-54$ for MGs TiZrHfBeCu and TiZrHfBeNi, respectively. It is to be emphasized that these value exactly equal to the $K$-values calculated above with Eq.(\ref{K}) using the shear susceptibilities $\beta_i$ and relaxation volumes $r_i$ verifying that these quantities were chosen correctly. 

On the other hand, provided that the shear susceptibility $\beta_i$ is known,  the relaxation volume $r_i$ of a defect can be calculated from  experimental data on the volume change  and temperature dependence of  corresponding variation of the defect concentration $\Delta c$ with respect to room temperature according to Eq.(\ref{DeltaV_V}).  The volume change of glass in this equation can be determined with Eq.(\ref{Omega}). Then, the dependence  $\Omega (T)$ vs $\Delta c (T)$ should constitute a straight line with a slope equal to $r_i-1$. This assumption is verified in Fig.5, which shows this dependence for the MGs under study in the initial and relaxed states. First of all it is seen that $\Omega(\Delta c)$-dependence  is indeed close to linear for both MGs. The dashed lines give the corresponding least square fits while their slopes $d\Omega/d\Delta c$ define the relaxation volumes $r_i$ equal to 1.45 and 1.43 for TiZrHfBeCu and TiZrHfBeNi, respectively. It is these $r_i$-values, which were accepted above for the calculations.      

\begin{figure}[t]
\begin{center}
\includegraphics[scale=0.68]{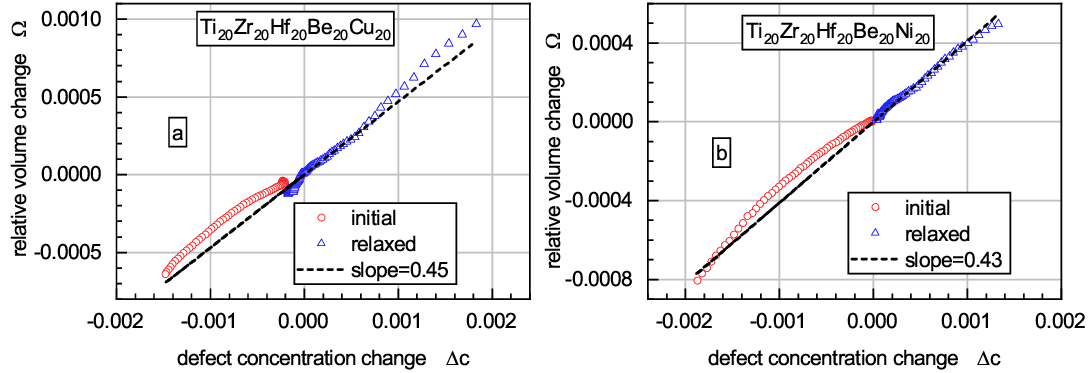}
\caption[*]{\label{Fig5.eps} Relative volume change $\Omega$ of glassy TiZrHfBeCu  (a) and TiZrHfBeNi (b) in the initial and relaxed states calculated using Eq.(\ref{Omega}) as a function  of the defect concentration change $\Delta c$  with respect to its room-temperature value determined with Eq.(\ref{Deltac}). The dashed lines give the least square fits. It is seen that  $\Omega (\Delta c)$-dependences are nearly linear and, therefore, the slopes  $d\Omega/d\Delta c$ provide the values of $r_i-1$,  as suggested by  Eq.(\ref{DeltaV_V}).  }
\end{center}
\end{figure}

Second, Fig.\ref{Fig5.eps} shows that if the defect concentration change $\Delta c\approx 0$ then the volume change $\Omega \approx 0$. A decrease of the defect concentration with respect to that at room temperature (i.e. $\Delta c<0$) due to structural relaxation below $T_g$ leads to a decrease of the volume. Such a situation takes place for initial samples (red circles in Fig.\ref{Fig5.eps}a,b). Conversely, an increase of the defect concentration over its room-temperature value (i.e. $\Delta c>0$) occurring upon heating of relaxed samples leads to an increase of  the volume (blue triangles). On the whole, the linear dependence $\Omega$ vs $\Delta c$ reflects the diaelastic effect  in MGs -- a decrease of the shear modulus with increasing defect concentration. 

It is important to emphasize also that the  \textit{dimensionless} functions $\delta (T)$, $\Omega (T)$ and $\Delta c(T)$  are obtained upon  scanning of the whole temperature range from room temperature up to the crystallization onset.  Each of the depedences, $\delta =K_i(\Omega)$  and $\Omega =f (\Delta c)$, shown in Fig.\ref{Fig4.eps} and Fig.\ref{Fig5.eps}, provides a unique characterization of a particular glass in both initial and relaxed states. Therefore, these dependences can be considered as master curves describing different aspects of the interrelation between the shear modulus and volume changes upon structural relaxation of glass both below and above $T_g$. The fact that  linear dependences $\delta (\Omega)$  and $\Omega (\Delta c)$ are governed by single paramaters $K_i$ and $r_i$, respectively, can be used for the prediction of $G$- and $\Delta V/V$-changes upon structural relaxation provided that these parameters are known. Meanwhile, since $K_i$ and $r_i$ can be easily estimated, MGs' relaxation behavior can be readily predicted as well.

\subsection{Comments on the relaxation mechanism }

Generally speaking, the slope $d\Omega/d\Delta c$ of the relative volume change $\Delta \Omega$  vs  defect concentration change $\Delta c$ (Fig.\ref{Fig5.eps}), which is found to be 0.43 to 0.45,  does not directly point at  interstitial-type defects as a source of relaxation. Alternatively, one suggest that the relaxation is conditioned by the free volume regions, which can be provisionally understood as vacancy-type defects. Accepting their properties to be close to those of vacancies, relative increase of the volume $\left(\Delta V/V\right)_v$ due to the generation of these defects with a concentration $\Delta c$ can be written down as \cite{GordonJNonCtystSol1996}

\begin{equation}
\left(\frac{\Delta V}{V}\right)_v=(r_v+1)\Delta c, \label{DeltaV_V_vac}
\end{equation}
where $r_v$ is the relaxation volume of a vacancy-type defect and $\Delta c$ is the defect concentration. Note that this equation is different from the analogous  equation (\ref{DeltaV_V}) for an interstitial-type defect. The differences between the volume changes produced by interstitial-type and vacancy-type defects are determined by their relaxation volumes, $r_i$ and $r_v$. The expected values for these quantities were first given by Granato \cite{GordonJNonCtystSol1996} as $r_v\approx -0.2$ and $r_i\approx 2.0$. Later studies found $-0.42\leq r_v\leq -0.05$  and $1.1\leq r_i\leq 1.9$ for different FCC and BCC metals (Al,Cu,Ni,Pt,$\alpha$-Fe,Mo) \cite{Ehrhart1991,Wolfer2012}, in general agreement with Granato's estimate. Simulation studies give similar values $r_v=-0.49$, $r_i=1.76$ for Al and $r_v=-0.36$, $r_i=1.57$ for high-entropy FeNiCrCoCu \cite{MakarovJPCM2020}. 

Thus, if one accepts a vacancy-type relaxation mechanism, then with the experimental value $d\Omega/d\Delta c \approx 0.45$, using Eq.(\ref{Omega}) and Eq.(\ref{DeltaV_V_vac}), one arrives at the relaxation volume of vacancy-type defects as $r_v\approx -0.55$, which is quite close to the aforementioned $r_v$ estimates.
On the other hand, if relaxation is conditioned by interstitial-type defects, one should use Eq.(\ref{DeltaV_V}) and Eq.(\ref{Omega}) for the relative volume change. Then, with  experimental  $d\Omega/d\Delta c \approx 0.45$ one arrives at the relaxation volume $r_i=1.45$, which is also fairly close to the experimental $r_i$-values indicated above. 

As mentioned above, the relation parameter $K_i$ (Eq.(\ref{K})) given above (equal to $-53$ and $-44$ for the MGs under study) is indicative of the interstitial-type relaxation. For vacancy-type relaxation, this parameter becomes \cite{MakarovJPCM2020}
\begin{equation}
 K_v=\left(\frac{dln\;G}{dln\;V}\right)_v=-\frac{\beta_v}{r_v+1}. \label{Kv}
 \end{equation} 
 Then, taking the estimates for the shear susceptibility $\beta_v=2$  and $r_v=-0.38$ \cite{MakarovJPCM2020} one arrives at $K_v\approx -3$. Thus, one should accept that  $\vert K_i\vert $ is by an order of magnitude larger than  $\vert K_v\vert $. 
 
Meanwhile,  volume changes induced by vacancy- and interstital-type defects are generally quite comparable. Indeed, if the same concentration $\Delta c$ of vacancy-type and interstitial-type defects is created then the ratio of the volume effects produced by these defects using Eqs(\ref{DeltaV_V}) and (\ref{DeltaV_V_vac}) can be represented as

\begin{equation}
\frac{\left(\Delta V/V\right)_i}{\left(\Delta V/V\right)_v}=\frac{r_i-1}{r_v+1}. \label{ratio_rel_vol}
\end{equation}
Then, accepting aforementioned $r_i=1.57$ and $r_v=-0.36$ found for high-entropy FeNiCrCoCu  \cite{MakarovJPCM2020} one can conclude that the ratio (\ref{ratio_rel_vol}) is about 0.9 and, therefore, interstitial-type and vacancy-type defects produce nearly the same volume effect. On the other hand, as mentioned above, the diaelastic effect produced by interstitial-type defects is by an order of magnitude larger and this leads to the aforementioned inequality   $\vert K_i\vert >> \vert K_v\vert$. 

In the view of the above, one can summarize that the relationship between shear modulus and volume relaxation in high-entropy MGs under study is conditioned by interstitial-type defects. The physical origin of this effect is fairly clear. It was long ago found that the atoms belonging to the nucleus of an interstitial defect are very mobile and display a few low-frequency vibration modes \cite{DederichsJNuclMater1978,Robrock1990}, which determine a large diaelastic effect. Low-frequency modes are also observed in the vibration spectra of MGs and these modes can be related to the interstitial-type defect system in these materials as suggested before \cite{KonchakovJPCM2019}. These modes provide additional anelastic strain under the external shear stress and lead to a large diaelastic effect, which is observed as a strong decrease of the shear modulus (shear softening) upon an increase of the defect concentration.        

Finally, it is to be  noted  that high-entropy TiZrHfBeCu and TiZrHfBeNi glasses under study have the parameter $K_i$, which is by the absolute magnitude either close ($K_i=-44$ for TiZrHfBeCu) or notably larger ($K_i=-53$ for TiZrHfBeNi) than $K_i$-values found earlier for conventional glasses \cite{MakarovJPCM2020}. This means that the diaelastic effect in high-entropy MGs can be larger as well. This in turn means that this  effect could be related to  \textit{low excess entropy}, which is characteristic of MGs with \textit{high mixing entropy} \cite{AfoninAPL2024}. Understanding of the role of the entropy in MGs' diaelasticity constitutes a challenging task for future research.

\section{Conclusions}
We performed parallel measurements of the high-frequency shear modulus $G$ and volume changes $\Delta V/V$ for high-entropy Ti$_{20}$Zr$_{20}$Hf$_{20}$Be$_{20}$Cu$_{20}$  and Ti$_{20}$Zr$_{20}$Hf$_{20}$Be$_{20}$Ni$_{20}$ glasses upon linear heating in the range from room temperature up to the complete crystallization. The changes of $G$ and $\Delta V/V$ occurring due to structural relaxation both below and above the glass transition temperature $T_g$ are singled out.  On this basis, the following conclusions are made.

1. Experimentally observed relaxation changes of the density (volume) are quite well described with Eq.(\ref{DeltaRho_Rho}), which is derived using the Interstitialcy theory and relates  relaxation changes of the shear  modulus with the concentration of interstitial-type defects frozen-in upon glass production as suggested by Eq.(\ref{Deltac}). This description is valid both below and above $T_g$ for both MGs under study, despite of significant differences in their relaxation behavior.  

2. It is shown that temperature-dependent parameter $\delta (T)$ defined by Eq.(\ref{delta_mu}) and reflecting temperature dependences of the shear moduli  the high-entropy glassy and crystalline states linearly depends on temperature-dependent relative volume change $\Omega (T)$ occurring upon relaxation and determined by relationship (\ref{Omega}). Meanwhile, the slope of this dependence $d \delta / d\Omega$ is nearly constant in the whole temperature range for both samples' states (initial/relaxed) and represents \textit{temperature-independent dimensionless } relaxation parameter $K_i=dln\;G/dln\;V$. This parameter equals to -44  for TiZrHfBeCu and to -53 for TiZrHfBeNi glass, respectively. These values by the absolute magnitude are either comparable or notably larger than those found earlier for conventional MGs. 

It is shown that the parameter $K_i$ is related to the shear susceptibility $\beta_i$ of glass and relaxation volume $r_i$ of defects responsible for relaxation as assumed by Eq.(\ref{K}). Independent estimate of $K_i$ using $\beta_i$- and $r_i$-values agrees very well with its determination from experimental $\delta (\Omega)$ dependence.

3. It is shown that temperature-dependent volume change $\Omega (T)$  linearly varies with temperature-dependent change of the defect concentration $\Delta c(T)$ occurring upon relaxation. The slope of this dependence, $d\Omega/d\Delta c$, constitutes another \textit{ dimensionless temperature-independent} relaxation parameter, which defines defects' relaxation volume $r_i$. It is found that $r_i$-values calculated using the  experimental data agree with the corresponding expected values for interstitial-type defects. 

4. The above dependences $\delta=K_i(\Omega)$ and $\Omega=f(\Delta c)$ are determined from experimental data in the whole temperature range of the experiment. In particular, these depedences show that \textit{i}) $\delta\approx 0$ at $\Omega = 0$ and \textit{ii}) $\Omega\approx 0$ at $\Delta c=0$ and these relations are valid for both initial and relaxed states of high-entropy MGs under investigation. Thus, these dependences constitute master curves, which describe the relaxation at different structural states and, besides that, can be used for the prediction of  the relaxation kinetics.

5. Possible contribution of vacancy-type defects into shear modulus vs volume relaxation is analyzed. It is concluded that, although vacancy- and interstitial-type defects produce nearly the same volume changes, the former defects cannot provide any significant diaelastic effect. All data obtained point quite convincingly at interstitial-type defects as a source of the relaxation.  

\section{Acknowledgments}

The work was supported by  Russian Science Foundation under the project 23-12-00162.


\begin{thebibliography}{999}

\bibitem{ChenRepProgPhys1980} H.S. Chen, Glassy metals, Rep. Prog. Phys. 43 (1980) 353-432.


\bibitem{Hirth} J.P. Hirth, J. Lothe, Theory of Dislocations (2nd ed.) Wiley, New York, 1982.

\bibitem{GranatoPRL1992} A.V. Granato, Interstitialcy model for condensed matter states of face-centered-cubic metals, Phys. Rev. Lett. 68 (1992) 974-977.

\bibitem{DyreRevModPhys2006} J.C. Dyre. The glass transition and elastic models of glass-forming liquids. Rev. Mod. Phys. 78 (2006) 953-972.

\bibitem{MakarovJNCS2019} A.S. Makarov, Yu.P. Mitrofanov, R.A. Konchakov, N.P. Kobelev, K. Csach, J.C. Qiao, V.A. Khonik. Density and shear modulus changes occurring upon structural relaxation and crystallization of Zr-based bulk metallic glasses: in situ measurements and their interpretation. J. Non-Cryst. Sol. 521 (2019) 119474. 

\bibitem{ChenJAP1978} H.S. Chen, The influence of structural relaxation on the density and Young's modulus of metallic glasses, J. Appl. Phys. 49 (1978) 3289-3291.

\bibitem{HarmsJNCS2003} U. Harms,O. Jin, R.B. Schwarz, Effects of plastic deformation on the elastic modulus and density of bulk amorphous Pd$_{40}$Ni$_{10}$Cu$_{30}$P$_{20}$, J. Non-Cryst. Sol. 317 (2003) 200-205.

\bibitem{ChengProgMaterSci2011} Y.Q. Cheng, E. Ma, Atomic-level structure and structure–property relationship in metallic glasses, Prog. Mater. Sci. 56 (2011) 379-473.

\bibitem{SpaepenActaMetall1977} F. Spaepen, A microscopic mechanism for steady state inhomogeneous flow in metallic glasses, Acta Metall. 25 (1977) 407-415.

\bibitem{ArgonActaMetall1979} A.S. Argon, Plastic deformation of metallic glasses, Acta Metall.  27 (1979) 47-58.

\bibitem{vdBeukelActaMetMater1990} A. van den Beukel, J. Sietsma, The glass transition as a free volume related kinetic phenomenon, Acta Met. Mater. 38 (1990) 383-389.

\bibitem{EgamiProgMaterSci2011} T. Egami, Atomic level stresses, Prog. Mater. Sci. 56 (2011) 637–653. 

\bibitem{LiuSciRep2017} S.T. Liu, F.X. Li,  M.Z. Li, W.H. Wang, Structural and dynamical characteristics of flow units in metallic glasses, Sci. Rep. 7 (2017) 11558.

\bibitem{KobelevUFN2023} N.P. Kobelev, V.A. Khonik, A novel view of the nature of formation of metallic glasses, their structural relaxation, and crystallization, Physics--Uspekhi 66 (2023) 673-690. 

\bibitem{MiracleMRSBull2007} D.B. Miracle, T. Egami, K.M. Flores, K.F. Kelton, Structural aspects of metallic glasses. MRS Bull. 32 (2007) 629-634.

\bibitem{ChengApplPhysLett2008} Y. Q. Cheng, E. Ma, Indicators of internal structural states for metallic glasses: Local order, free volume, and configurational potential energy, Appl. Phys. Lett. 93, 051910 (2008).

\bibitem{HaruyamaApplPhysLett2006} O. Haruyama, A. Inous, Free volume kinetics during sub-Tg structural relaxation of a bulk Pd$_{40}$Ni$_{40}$P$_{20}$ metallic glass, Appl. Phys. Lett. 88 (2006) 131906. 


\bibitem{HaruyamaActaMater2010} O. Haruyama, Y. Nakayama, R. Wada, H. Tokunaga, J. Okada, T. Ishikawa, Y. Yokoyama, Volume and enthalpy relaxation in Zr$_{55}$Cu$_{30}$Ni$_{5}$Al$_{10}$ bulk metallic glass, Acta Mater. 58 (2010) 1829-1836.


\bibitem{EvensonActaMater2011} Z. Evenson, R. Busch, Equilibrium viscosity, enthalpy recovery and free volume relaxation in a Zr$_{44}$Ti$_{11}$Ni$_{10}$Cu$_{10}$Be$_{25}$ bulk metallic glass, Acta Mater 59 (2011) 4404-4415.

\bibitem{BunzJAP2013} J. B\"{u}nz, G. Wilde, Direct measurement of the kinetics of volume and enthalpy relaxation of an Au-based bulk metallic glass, J. Appl. Phys. 114 (2013) 223503.

\bibitem{MakarovJPCM2020} A.S. Makarov, R.A. Konchakov, Yu.P. Mitrofanov, M.A. Kretova, N.P.  Kobelev, V.A. Khonik, A simple kinetic parameter indicating the origin of the relaxations induced by point(-like) defects in metallic crystals and glasses, J. Phys.: Condens. Matter 32 (2020) 495701.

\bibitem{DuEncMater2022} Y. Du, Q. Zhou, H. Wang. High-entropy alloys: bulk metallic glasses. Encyclopedia of materials: metals and alloys, 2 (2022) 318-326.

\bibitem{DederichsJNuclMater1978}  P.H. Dederichs, C. Lehman, H.R. Schober, A. Scholz, R. Zeller, Lattice theory of point defects, J. Nucl. Mater. 69-70 (1978) 176-199. 

\bibitem{HolderPhysRevLett1974} J. Holder,  A.V. Granato, L.E. Rehn,  Experimental evidence for split interstitials in copper, Phys. Rev. Lett 32 (1974)  1054-1057.

\bibitem{RobrockRadEff1975}  K.-H. Robrock, V. Spiric, L.E. Rehn, Measurement of the diaelastic effect in polycrystalline copper during electron irradiation at 5 K, Radiation Effects 27 (1975) 189-190.

\bibitem{JiangNatCommun2021} J. Jiang, Z. Lu, J. Shen, T. Wada, H. Kato, M. Chen, Decoupling between calorimetric and dynamical glass transitions in high-entropy metallic glasses, Nat. Commun. 12 (2021) 3843.

\bibitem{DuanPRL2022} Y.J. Duan, L.T. Zhang, J.C. Qiao, Y.-J. Wang, Y. Yang, T. Wada, H. Kato, J.M. Pelletier, E. Pineda, D. Crespo, Intrinsic correlation between the fraction of liquidlike zones and the $\beta$ relaxation in high-entropy metallic glasses, Phys. Rev. Lett. 129 (2022) 175501. 

\bibitem{LuanNatComm2022}  H. Luan, X. Zhang, H. Ding, F. Zhang, J.H. Luan, Z.B. Jiao, Y.-C. Yang, H. Bu, R. Wang, J. Gu, C. Shao, Q. Yu, Y. Shao, Q. Zeng, N. Chen, C.T. Liu, K.-F. Yao, High-entropy induced a glass-to-glass transition in a metallic glass, Nat. Comm. 13 (2022) 2183.

\bibitem{DuanPRL2024} Y.J. Duan, M. Nabahat, Y. Tong, L. Ortiz-Membrado, E. Jim\'{e}nez-Piqu\'{e}, K. Zhao, Y.-J. Wang, Y. Yang, T. Wada, H. Kato, J.M. Pelletier, J.C. Qiao, E. Pineda, Phys.Rev.Lett. 132 (2024) 056101.

\bibitem{AfoninAPL2024} G.V. Afonin, J.C. Qiao, A.S. Makarov, R.A. Konchakov, E.V. Goncharova, N.P. Kobelev, V.A. Khonik, high entropy metallic glasses, what does it mean? Appl. Phys. Lett. 124 (2024) 151905.

 
\bibitem{Vasil’evUFN1983} A.N. Vasil’ev, Yu. P. Gaidukov, Electromagnetic excitation of sound in metals, \textit{Soviet Physics Uspekhi} 26 (1983) 952-973.
 
\bibitem{AfoninJNCS2017} G.V. Afonin, Yu.P. Mitrofanov, A.S. Makarov, N.P. Kobelev, V.A. Khonik,  On the origin of heat effects and shear modulus changes upon structural relaxation and crystallization of metallic glasses, J. Non-Cryst. Sol. 475 (2017) 48-52.
 
\bibitem{AfoninJALCOM2024} G.V. Afonin, J.C. Qiao, A.S. Makarov, N.P. Kobelev, V.A. Khonik, Fast relaxation in metallic glasses studied by measurements of the internal friction at high frequencies, J. Alloys Comp. 996 (2024) 174783. 
 
\bibitem{GordonJNonCtystSol1996} C.A. Gordon, A.V. Granato, R.O. Simmons, Evidence for the self-interstitial model of liquid and amorphous states from lattice parameter measurements in krypton, J. Non-Cryst. Sol. 205-207 (1996) 216-220. 

\bibitem{Ehrhart1991} P. Ehrhart, Properties and interactions of atomic defects in metals and alloys, in: Landolt-Bornstein New Series III, Vol. 25, ed. O. Madelung (Springer, Berlin, 1991) p. 88-371.
 
\bibitem{Wolfer2012} W.G. Wolfer. Fundamental properties of defects in metals, in: Comprehensive Nuclear Materials (Ed. R.J.M. Konings), Elsevier: NewYork, NY, USA, 2012, p.1-45.


\bibitem{Robrock1990} K.H. Robrock. Mechanical Relaxation of Interstitials in Irradiated Metals. Berlin, Springer-Verlag,  1990.

\bibitem{KonchakovJPCM2019}	R.A. Konchakov, A.S. Makarov, N.P. Kobelev, A.M. Glezer, G. Wilde, V.A. Khonik. Interstitial clustering in metallic systems as a source for the formation of the icosahedral matrix and defects in the glassy state. J. Phys.: Cond. Matter 31 (2019) 385703.

\end{thebibliography}
\end{document}